\documentclass{aa}
\usepackage{graphics}

\newlength{\figwidth}
\begin{document}
\title{Helioseismic limit on heavy element abundance}
\titlerunning{Helioseismic limit on heavy element abundance}
\author{H. M. Antia \inst{1} \and S. M. Chitre \inst{2}}
\authorrunning{Antia \and Chitre}
\offprints{H. M. Antia, \email{antia@tifr.res.in}}
\institute{Tata Institute of Fundamental Research,
Homi Bhabha Road, Mumbai 400005, India
\and
Department of Physics, University of Mumbai, Mumbai 400098,
India}
\date{Received }

\abstract{
Primary inversions of accurately measured solar oscillation frequencies coupled
with the equations of thermal equilibrium and other input physics,
enable us to
infer the temperature and hydrogen abundance profiles inside the Sun.
These profiles also help in setting constraints on the input physics
as well as on heavy element abundance in the solar core.
Using different treatments of plasma screening for nuclear reaction
rates, limits
on the cross-section of proton-proton nuclear reaction as
a function of heavy element abundance in the
solar core are obtained and an upper limit on heavy element
abundance in the solar core is also derived from these results.}

\maketitle

\section{Introduction}

The precisely measured frequencies of solar oscillations have been used
to probe the solar interior.
The primary inversions of these observed frequencies yield 
the sound speed and density profiles inside the Sun.
In order to infer the temperature and chemical
composition profiles, we also need to know the input
physics such as opacities, equation of state and nuclear energy
generation rates (Gough \& Kosovichev \cite{dog88};
Kosovichev \cite{kos96};
Shibahashi \& Takata~\cite{st96}; Takata \& Shibahashi~\cite{tak98};
Antia \& Chitre \cite{ac98}). In all these works the heavy element
abundance profile is assumed to be known; attempts to determine
heavy element abundance profile from helioseismic data have not
been particularly successful (Antia \& Chitre \cite{ac99};
Takata \& Shibahashi \cite{tak01}) as the resulting inverse problem becomes
extremely ill-conditioned.
Fukugita \& Hata~(\cite{fuk98}) have obtained limits on heavy
element abundance in the solar core using observed solar neutrino fluxes.
It would be interesting to enquire if such limits can be independently obtained
from helioseismic data.

In general, the computed luminosity in a seismically computed solar
model is not expected to match the observed solar luminosity.
By applying
the observed luminosity constraint it is possible to constrain the
input physics, particularly, the
cross-section of proton-proton (pp) nuclear reaction.
Antia \& Chitre (\cite{ac98})
estimated this cross-section to be
$S_{11}= (4.15\pm0.25)\times10^{-25}$ MeV barns.
Similar values have been obtained by comparing the computed solar
models with helioseismic data
(Degl'Innocenti, Fiorentini \& Ricci~\cite{inn98};
Schlattl, Bonanno \& Paterno~\cite{bon99}).
The main source of error in these
estimates is the uncertainty in the $Z$ profile and, therefore, Antia \& Chitre
(\cite{ac99}) attempted to find the pp reaction rate as a function
of $Z$ in the solar core. In all these works the plasma screening
of nuclear reaction cross-sections was calculated using intermediate
screening formulation of Graboske et al.~(\cite{gra73}). The treatment of 
screening in stellar nuclear reaction rates is not yet adequately
understood (Dzitko et al.~\cite{dzi95};
Gruzinov \& Bahcall \cite{gru98}).
Wilets et al.~(\cite{wil00}) 
have done a sophisticated treatment of plasma screening
and compared their results with earlier prescriptions.
Their results indicate that for the solar core the intermediate
screening treatment due to Mitler (\cite{mit77}) is better than
that due to Graboske et al.~(\cite{gra73}).
It would thus be interesting to study the effect of different
treatment of plasma screening on the helioseismically
estimated pp reaction cross-section.

Antia \& Chitre (\cite{ac99}) included the effect of heavy element
abundance $Z$ only on the opacity of the solar material. If we
make the reasonable assumption
that the abundances of C, N, O also increase with $Z$, then the
CNO cycle will become more effective in contributing to the
nuclear energy generation in the
solar core. At normally accepted values of $Z$ it is estimated that less than
2\% of energy generated in the central region is produced by the
CNO cycle (Bahcall et al.~\cite{bp01}).
But if $Z$ value is increased, this proportion will clearly increase and
consequently, the pp reaction rate needs to be reduced to maintain
the observed solar luminosity. In this work we demonstrate that this effect
can be exploited to set an upper limit
on $Z$ in the solar core.

\section{The technique}

With the use of accurately measured p-mode frequencies from the first
year of operation of MDI instrument (Rhodes et al.~\cite{mdi97}) we infer
the sound speed and density profiles adopting a Regularised
Least Squares technique (Antia \cite{a96}). The determination of thermal
and chemical composition profiles, however, necessitates the use
of equations of thermal equilibrium (Antia \& Chitre \cite{ac98}) with the
supplementary input of the equation of state, opacities
and nuclear energy generation rates, provided we have a knowledge
of the heavy element abundance profile inside the Sun.
There is no guarantee for the resultant seismic model to yield the
observed solar luminosity, $L_\odot=3.846\times 10^{33}$ ergs s$^{-1}$,
unless we adjust the nuclear reaction rates slightly. 
This turns out to provide a valuable tool for constraining the
nuclear reaction rates, in particular, the cross-section of the
pp nuclear reaction which has not been measured in the laboratory
because of the slow weak interaction rate. The pp reaction cross-section
has only been calculated theoretically and it would be instructive
to test the validity of calculated results using the helioseismic
constraints.

There is an uncertainty of about 2\% in evaluating the luminosity
of seismic models because of possible errors in primary inversions
(including an estimate of systematic errors), solar radius, equation of state,
opacity, nuclear reaction rates for other reactions.
Possible error due to uncertainties in treatment of plasma screening
of nuclear reactions is not included in this error estimate.
This can be estimated by using different treatments of plasma screening.
The uncertainty arising from errors in $Z$ profiles is, however,
much larger (cf., Antia \& Chitre \cite{ac98}). We, therefore, estimate
for each $Z$ profile, the range of cross-section of pp nuclear
reaction, which reproduces the luminosity to within 2\% of the
observed value. The computed luminosity in the seismic models
will, of course, depend on the metallicity, $Z_c$ in the solar
core; we therefore, attempt to delineate the region  in the
$S_{11}$-$Z_c$ plane which gives the correct solar luminosity.
In order to obtain the thermal structure we adopt the OPAL
opacities (Iglesias \& Rogers \cite{igl96}),
the OPAL equation of state (Rogers, Swenson \& Iglesias \cite{rog96})
and the nuclear reaction rates from
Adelberger et al.~(\cite{fusion}).
The plasma screening
effects are calculated using either weak screening
(Salpeter \cite{sal54}) or intermediate screening (Mitler \cite{mit77}).
The effect of variation of $Z_c$ on nuclear energy generation rate is
incorporated in the computations by assuming the abundances of all
heavy elements to increase in the same ratio for obtaining the
abundances of C, N, O.

\section{Results}

\begin{figure}
\resizebox{\figwidth}{!}{\includegraphics{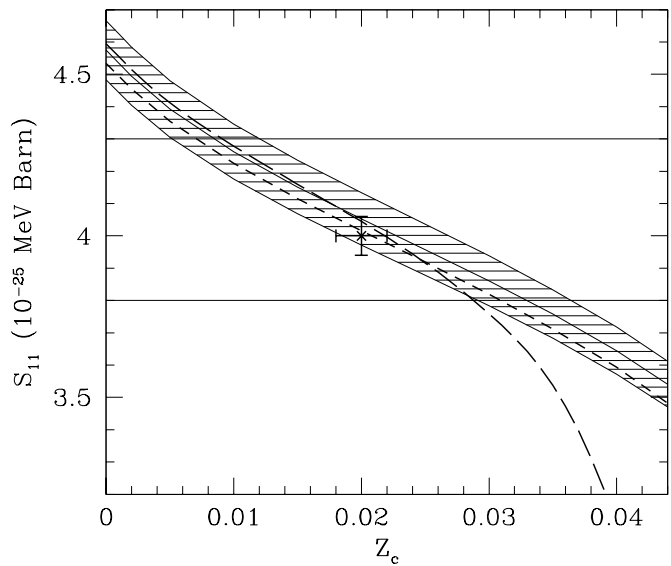}}
\caption{
The region in $Z_c$--$S_{11}$ plane that is consistent with helioseismic
data is marked by horizontal shading.
The continuous line defines the values obtained using intermediate
plasma screening, where
the seismic model matches the observed solar luminosity.
The point
with error bars shows the current best estimates for $Z_c$ and
$S_{11}$. The short-dashed line shows the central value obtained using
weak screening, while the long-dashed line shows the central value
obtained using intermediate screening and also including the
effect of $Z$ on nuclear energy generation. The two horizontal lines
mark the limits of theoretically calculated values of $S_{11}$.
}
\label{f1}
\end{figure}

With the help of the inverted profiles for sound speed and density,
along with a homogeneous $Z$ profile, covering a wide range of $Z$
values,
we first calculate a seismic model
by employing the equations of thermal equilibrium.
For each central value of $Z$ we estimate the range of
cross-section of pp nuclear reaction, which reproduces the luminosity
to within 2\% of the observed value. The numerical results are shown in Fig.~1,
which delineates the region in $Z_c$--$S_{11}$ plane that is
consistent with helioseismic and luminosity constraints. This figure
shows the results obtained using intermediate screening (Mitler \cite{mit77}).
For comparison the central value obtained using weak screening is also
shown in this figure. In order to compare these results with the earlier
work of Antia \& Chitre (\cite{ac99}), in these calculations the
effect of $Z$ on nuclear energy generation rate is not included.
Thus the difference is mainly due to treatment of plasma screening.
It is clear that with these treatments of screening the
current estimate of pp reaction cross-section (Adelberger et al.~\cite{fusion})
is consistent with helioseismic estimate within the expected error bars,
though the best value is about 1.5\% higher than the theoretical estimates.
Further, there is not much difference between results obtained using
the two different treatments of plasma screening. In general, the
estimated value of $S_{11}$ using weak screening is smaller than that
obtained using intermediate screening by about 1\%.

\begin{figure}
\resizebox{\figwidth}{!}{\includegraphics{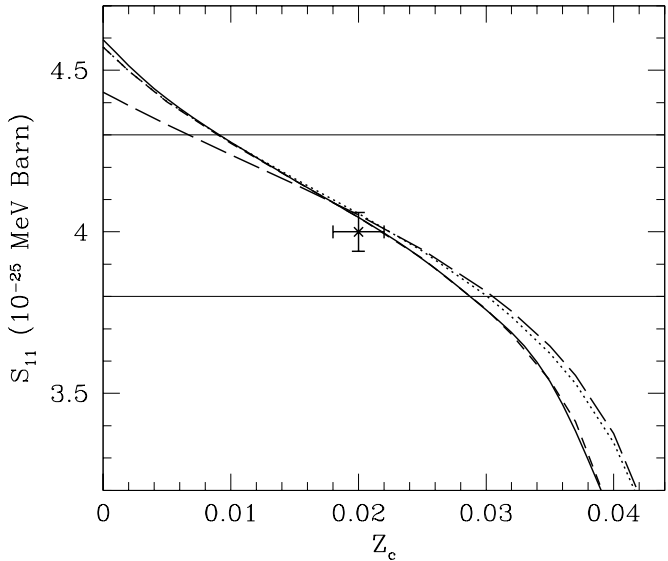}}
\caption{
The seismically estimated value of $S_{11}$ as a function of
$Z_c$ the heavy element abundance at the centre for different
$Z$ profiles. The solid line shows the value obtained with
homogeneous $Z$ profile and is the same as the long-dashed line in
Fig.~\ref{f1}. The short-dashed and long-dashed lines show the
results obtained using $Z$ profile given by Eq.~\ref{e1} and \ref{e2},
respectively. The dotted line shows the results using $Z$ profile
of Eq.~\ref{e1},
but with the reaction rate of $^{14}{\rm N}+p$ reaction reduced by
40\%.
The point
with error bars shows the current best estimates for $Z_c$ and
$S_{11}$. 
}
\label{f2}
\end{figure}

With a view to study the effect of $Z$ on energy generation through CNO cycle,
we repeat the calculations by varying the abundances of all heavy elements
in the same proportion as $Z$ in
the nuclear reaction network. These results are
shown by long-dashed line in Fig.~1. We do not expect
much difference at low values of $Z_c$, since CNO cycle produces a
tiny fraction of energy in the Sun. But with augmented $Z_c$, this fraction
becomes larger due to increased $Z$ as well as due to the
resulting increase in the temperature,
since the CNO reactions are more sensitive to temperature.
As a consequence, in order to maintain the solar luminosity constraint we need to
reduce $S_{11}$, with the result that around $Z_c= 0.035$ the estimated value of
$S_{11}$ decreases very sharply. This is because a significant fraction of the luminosity is
then accounted for by the CNO reactions with seismically inferred
temperature and composition profiles. For $Z_c=0.035$, CNO cycle
accounts for about 8\% of solar luminosity and this fraction,
in fact, increases rapidly with $Z_c$.
Thus, we can consider this
as the upper limit on $Z_c$, which is somewhat larger than
that inferred by Fukugita and Hata (\cite{fuk98}). A lower
limit on $Z_c$ can also be obtained from this figure if we
accept the range of theoretically computed values of $S_{11}$
(Bahcall \& Pinsonneault \cite{bp95}; Turck-Chi\'eze \& Lopes \cite{tc93})
as representing the acceptable range of $S_{11}$. Assuming an
upper limit of $4.3\times10^{-25}$ MeV barns for
$S_{11}$ we get a lower limit of $Z_c=0.006$, which is again
comparable to the lower limit obtained by Fukugita and Hata (\cite{fuk98}).
If we adopt a lower limit of $3.8\times10^{-25}$ MeV barns for $S_{11}$,
the resulting upper limit on $Z_c=0.032$. The exact limits on $Z_c$
will of course, depend on the assumed limits on $S_{11}$, but because
of the sharp fall in the long-dashed curve in Fig.~1, the upper limit
is not particularly sensitive to the assumed lower limit on $S_{11}$
and any value $Z_c>0.035$ would certainly be difficult to reconcile with
seismic models.

In the foregoing calculations we have adopted a homogeneous $Z$ profile,
which is perhaps not too realistic.
Since most of the energy generation occurs in the solar core ($r\le0.25R_\odot$)
our results are unaffected by the $Z$ profile in the outer radiative
region. To demonstrate this we also use the following $Z$ profiles :
\begin{equation}
Z=\cases{Z_c& if $r<0.25R_\odot$,\cr
Z_c+{(Z_s-Z_c)\over 0.463}(r/R_\odot-0.25)& otherwise,\cr}\label{e1}
\end{equation}
and
\begin{equation}
Z= Z_c+(Z_s-Z_c)r/(0.713R_\odot),\label{e2}
\end{equation}
where $Z_s=0.018$ is the known value of $Z$ at the solar surface.
The seismically estimated value of $S_{11}$ for these $Z$ profiles
is compared with those for the homogeneous profile in Fig.~\ref{f2}.
As to be expected, for $Z$ profile given by Eq.~\ref{f1} the results are
almost identical to those obtained with the homogeneous profile, while for the
linear $Z$ profile given by Eq.~\ref{e2} the average value of $Z$
in the solar core would be somewhat lower and as a result the estimated
$S_{11}$ values are somewhat larger, but the difference is comparable
to the estimated errors. If instead of central value of $Z$ we had
used the value of $Z$ at around $0.1R_\odot$ for the linear profile,
the two curves would have been almost identical.
Thus it is clear that exact form of the $Z$
profile is not very important and the upper limit on $Z_c$ is only
weakly dependent on the $Z$ profile.

Apart from uncertainties in $Z$ profile the nuclear reaction rate
for the CNO cycle reactions are also rather uncertain and these may
affect the inferred upper limit. To estimate
this uncertainty we repeat the calculations using the profile given
by Eq.~\ref{e1}
for $Z$, but with the reaction rate of the $^{14}{\rm N}+p$ reaction
reduced by 40\%, which is the estimated uncertainty in this reaction
rate (Adelberger et al.~\cite{fusion}). These results are also shown by the dotted
line in Fig.~\ref{f2}. From the figure it is clear the uncertainties
in CNO reaction rates will not particularly affect the upper limit on $Z_c$
in a significant way.

\begin{table}[t]
\begin{center}
\caption{Neutrino fluxes in a seismic model using the $Z$ profile
from model N0 of Brun et al.~(\cite{bru02}).}
\begin{tabular}{ll}
\hline
Source&Flux (cm$^{-2}$ s$^{-1}$)\\
\hline
pp&$(6.05\pm0.06)\times10^{10}$\\
pep&$(1.42\pm0.02)\times10^8$\\
hep&$2.09\times10^3$\\
$^7$Be&$(4.76\pm0.48)\times10^9$\\
$^8$B&$(4.83\pm0.88)\times10^6$\\
$^{13}$N&$(5.27\pm0.79)\times10^8$\\
$^{15}$O&$(4.48\pm0.80)\times10^8$\\
$^{17}$F&$(2.76\pm0.30)\times10^6$\\
Total Cl&$7.27\pm1.14$ SNU\\
Total Ga&$127.8\pm7.2$ SNU\\
\hline
\end{tabular}
\end{center}
\end{table}

Table~1 lists the neutrino fluxes in a seismic model obtained using
the Z-profile of a model including tachocline mixing of
Brun et al.~(\cite{bru02}), with the use of intermediate screening,
for calculating the nuclear energy generation rates.
In this case we need to increase
$S_{11}$ by 1.6\% over the currently accepted value. As noted by
Brun et al.~(\cite{bru02}), an increase in $S_{11}$ leads to
a better agreement between solar models and the seismically inferred
sound speed and density profiles. Thus a small increase in $S_{11}$
may still be required, but it is within the errors in
helioseismic estimates and those in theoretically computed values.
These neutrino fluxes are generally larger than those in seismic model
of Antia \& Chitre (\cite{ac98}), presumably because of reduction in
$S_{11}$.
The $^8$B neutrino flux in the seismic model
is consistent with the current estimate of $(5.09\pm0.65)\times10^6$
cm$^{-2}$ s$^{-1}$ using the neutral current channel of
Sudbury Neutrino Observatory (SNO)
(Ahmad et al.~\cite{sno02}).

\section{Discussion and Conclusions}

With the help of inverted sound speed and density profiles,
it is possible to infer
the $T,X$ profiles in the solar interior, provided the $Z$ profile and
the input physics are known.
The resulting seismic models have the correct solar
luminosity, provided the heavy element abundance $Z_c$ in the solar core and
the cross-section for pp nuclear reaction are within the
shaded region shown in Fig.~1. It appears that the
currently accepted values of $Z_c$
or $S_{11}$ need to be increased marginally to make them consistent
with helioseismic constraints. The required increase is within the
error estimates. The higher estimates for $S_{11}$ obtained earlier
were due to differences in treatment of plasma screening.
With the use of weak (Salpeter \cite{sal54}) or intermediate screening due to Mitler (\cite{mit77})
the theoretically estimated value of $S_{11}$ is in reasonable
agreement with seismically estimated value. With a $Z$ profile in
a standard solar model N0 of Brun et al.~(\cite{bru02}), the seismically
estimated value of $S_{11}$ is $4.07\times10^{-25}$ MeV barns
with the use of intermediate screening and $4.02\times10^{-25}$ MeV barns
for weak screening used while calculating the nuclear energy
generation rate.

If the value of  heavy element abundance in the solar core, $Z_c$,
is increased beyond 0.035 the CNO cycle generates
a good fraction of solar luminosity and it is not possible to get
any consistent seismic model unless $S_{11}$ is decreased
substantially below the accepted value. This puts a clear upper limit
on the heavy element abundance in the solar core, which is
comparable to that independently obtained by Fukugita and Hata (\cite{fuk98}).
This upper limit is not very sensitive to the $Z$ profile or to
the uncertainties in the CNO reaction rates. Even if the additional
heavy elements in solar core do not include CNO, the effect of
opacity alone will also put an upper limit on $Z_c$, but in that
case the limit will depend on the assumed lower limit on $S_{11}$.
For currently accepted theoretical limits, the upper limit in this
case turns out to be around 0.035.

Note that the
seismic model satisfies the normal stellar structure equations,
though the inferred $X$ profile may not match that given by an evolutionary
solar model. Further, since the seismic model is confined to the
radiative interior, it will not be possible to match it to an acceptable
convection zone model, unless the heavy element abundance
at the top of the radiative region is close to the known surface value.
Thus a relatively high value of $Z$ in the solar core can be realised only
if the $Z$ profile has a significant gradient in the radiative region.
From experiments with varying $Z$ or $S_{11}$ in evolutionary solar
models also it is known that an increase in $Z$ can be compensated
for by a reduction in $S_{11}$ to match the seismically inferred
sound speed and density profiles (Brun et al.~\cite{bru02}). Thus it is
quite possible that a comparable upper limit on $Z_c$ may be obtained
from these models if we restrict the range of $S_{11}$, provided the
initial $Z$ profile at  the zero age main sequence stage is not
homogeneous.

\begin{acknowledgements}
SMC is grateful to DAE-BRNS for support under the Senior
Research Scientist Scheme
and to Ian Roxburgh for supporting his visit to Queen Mary, University
of London under the Leverhulme Trust Visiting Professorship scheme.
We thank the Referee, S. Degl'Innocenti for valuable suggestions,
which have led to an improvement in the presentation of our results.
\end{acknowledgements}

\end{document}